\documentclass[a4paper]{article}
\usepackage[english]{babel}
\usepackage{amsmath,amssymb}
\usepackage{cite,./mcite}
\usepackage{xspace}
\usepackage{endnotes}
\usepackage{epsfig}
\usepackage{rotate}
\usepackage{./sidecaption}
\usepackage{./tdrdefinitions}
\begin{document}

\def\abstracts#1#2#3{{
        \centering{\begin{minipage}{4.5in}\baselineskip=10pt\footnotesize
        \parindent=0pt #1\par 
        \parindent=15pt #2\par
        \parindent=15pt #3
        \end{minipage}}\par}} 

\mbox{ } \hfill   {\footnotesize BONN-HE-2001-01}\\
\mbox{ } \hfill   {\footnotesize June, 2001}\\

\vskip 8mm
\centerline{\bf Diffractive Production of Neutral Vector Mesons at THERA}
\vspace*{0.15truein}
\centerline{\footnotesize James A. Crittenden
\footnote{Present address: Deutsches Elektronen-Synchrotron, Notkestr. 85, 22603 Hamburg, Germany}
}
\vspace*{0.015truein}
\centerline{\footnotesize\it Physikalisches Institut, University of Bonn, Nu{\ss}allee 12}
\baselineskip=10pt
\centerline{\footnotesize\it 53115 Bonn, Germany}
\vspace*{10pt}

\vskip 8mm

\abstracts{
We consider the contribution to our understanding of 
vacuum-exchange processes to be made by 
investigations at the proposed electron-proton col\-lider THERA.
Recent results have highlighted the value of such studies for testing
quantum chromodynamical descriptions of both long-range and short-range
strong interactions. Stringent quantitative constraints
have been provided by
exploiting the opportunity to correlate scaling behaviour with
helicity selection in exclusive and semi-exclusive vector-meson production.
After reviewing the progress achieved by the measurement programs presently
being carried out by the H1 and ZEUS collaborations at HERA, we
discuss the performance criteria imposed by such investigations on 
the THERA accelerator complex and on the detector design. 
We conclude that the study of vector-meson production will form
an essential component of the THERA physics program
beginning with the early 
turn-on stage of the machine and continuing throughout
the achievement of its full high-luminosity potential.
}{}{}

\section{Introduction}

Investigations of vector-meson production at THERA will confront 
our understanding of strong interaction dynamics, meson and
baryon partonic structure, confinement mechanisms, flavour symmetries, 
scaling laws and helicity selection rules
with detailed and multi-various extensions of the wealth of information
obtained  from the HERA programs presently being
carried out by the H1 and ZEUS collaborations.
Along with an unprecedented ability for detailed investigations of elastic
and total photon-proton cross sections in the Regge limit of high
energy, the electron-proton collider experimental strategy  
has established the research field
of short-distance vacuum-exchange processes, or ``hard diffraction'',
as a subject of essential importance to studies of Quantum
Chromodynamics.
While many of these research 
topics have built on a foundation of knowledge derived from
decades of measurements, many others are completely new and only beginning
to receive theoretical attention.
All the topics have stimulated
widespread theoretical and phenomenological interest, confirming
existing theoretical prejudices in some cases, and clearly guiding theoretical
approaches in others. The experimental opportunity presented by the THERA
accelerator design directly addresses the need for an extension of the
energy range both for approaches based on Regge phenomenology
and for those based on perturbative and nonperturbative QCD. 
Even more importantly,
THERA will extend the kinematic reach in momentum transfer by more than an
order of magnitude, providing clean and detailed test of 
quantitative perturbative calculations of strong vacuum-exchange processes.

\begin{figure}
\sidecaption
  \epsfig{file=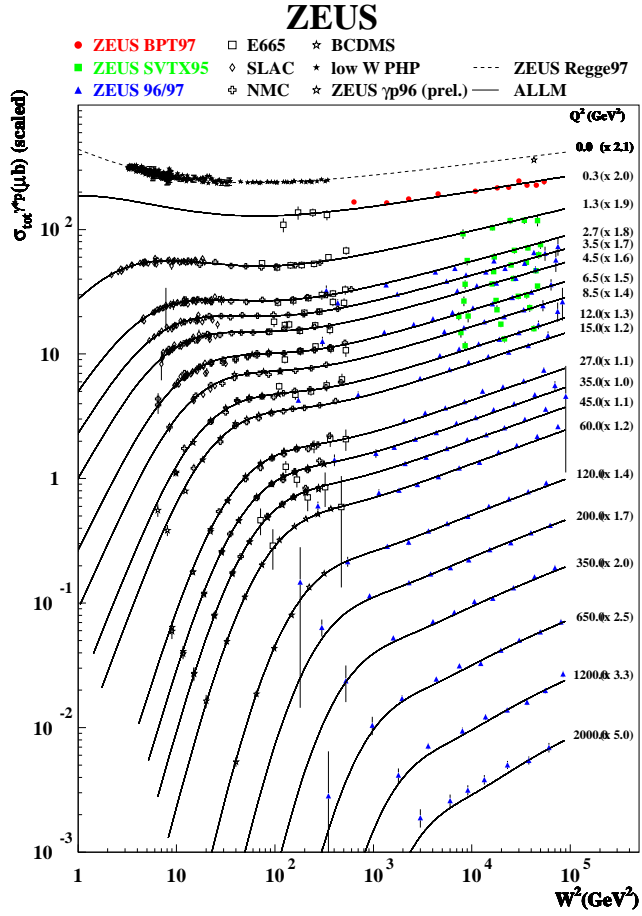,width=0.6\textwidth}
  \caption{
The
photon--proton cross section $\sigma^{{\gamma^*p}}_{tot}$ as a function of
the squared centre--of--mass energy for various values of Q$^2$.
The curves represent
calculations using the ALLM parton
distribution function parameterisations~\protect\cite{pl_269_465}
}
  \label{fig:sig}
\end{figure}

There are several simple reasons for the extraordinary variety of
theoretical physics concepts addressed by the study of
diffractive vector-meson
production in electron-proton interactions. Even at a fixed electron-proton
centre-of-mass energy, a broad range of energies
in the photon-proton centre-of-mass system is available
for investigation. Such a broad range of energy is of essential importance
for studies of
the weak energy dependence of soft diffractive processes.  The
corresponding access to the low-$x$ region means that
the coherence length of the virtual photons in the proton rest frame 
is much longer than the diameter of the proton~\cite{hardproc}, 
resulting in an unambiguous
definition of virtual-photon/proton cross sections and hence the opportunity
to study their scaling behaviour and helicity-transfer characteristics. 
Figure~\ref{fig:sig} shows the energy dependence of this cross section
measured at HERA.

\begin{figure}
\begin{minipage}{\textwidth}
\begin{center}
  \epsfig{file=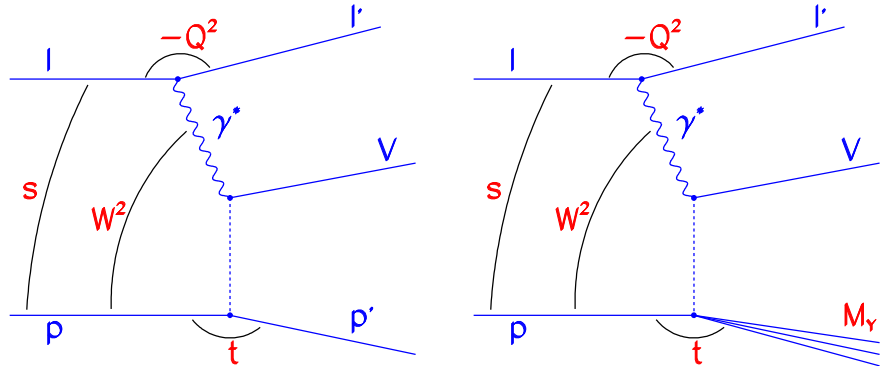,width=0.85\textwidth}
\end{center}
\end{minipage}
  \caption{
Schematic diagrams of a)~exclusive and b)~semi-exclusive electroproduction
of vector mesons. Such processes permit the study of vacuum-exchange
processes in perturbative and nonperturbative kinematic domains, including
the correlation of the helicity-transfer structure with the observed
power-law scaling with momentum transfer for momentum transfers exceeding
the hadronic confinement scale. The proton-dissociative process can
be studied in both the nucleon-resonance and high-mass regions, providing
information on Regge factorisation and on proton structure
}
  \label{fig:vmprod}
\end{figure}

The remarkably steep energy dependence at high photon
virtuality reflects the steep rise in the F$_2$ structure function at low~$x$.
Such
high energies and wide rapidity ranges provide access to the kinematic 
region of diffractive processes, where the momentum transfers are 
much
smaller than the kinematic limit. 
In the context of Regge phenomenology, recent investigations interpret 
the Q$^2$ dependence exhibited in
the photon-proton cross section
as evidence for the discovery of a second, ``hard'' 
Pomeron~\cite{hep-ph/0105088}. 
HERA results also 
show that accurate measurements can be made at momentum
transfers far exceeding the hadronic confinement 
scale yet also fulfilling this diffractive condition.
The analysis of exclusive and semi-exclusive vector-meson
production (see Fig.~\ref{fig:vmprod}) in 
the photon-dissociation region ensures the selection of
vacuum-exchange processes~\cite{jacjlab}. 
Since this selection can be done with little influence on the 
kinematics of the reaction, general questions concerning its characteristics
can be addressed, in particular the momentum-transfer scaling behaviour.
At low momentum transfer, this permits
investigation of phenomena described by Regge theory, such as 
the Pomeron trajectory and unitarity~\cite{kovchegov}. 
At high momentum
transfer, such experimental studies address the strong interaction dynamics
of vacuum exchange on distance scales much smaller than the
confinement scale, presently a subject of active theoretical speculation.
Particular to these studies of the production of 
phase-space-isolated vector mesons
is the clean experimental environment allowing exploitation of the
two-charged-particle decays to measure spin-density matrices with high
accuracy. Quantitative assessment of helicity-violating amplitudes
provides information on the partonic structure of the vector meson~\cite{pr_58_114026,*np_545_505,*jetpl_69_294}.  
Since the momentum-transfer scaling behaviour is
correlated to the helicity structure
of the interaction~\cite{pr_22_2157,*prep_112_173,pr_56_2982},
these measurements provide strict constraints on field theoretical approaches
and the associated power-law scaling features.
The proton-dissociative process can
be studied in both the nucleon-resonance and high-mass regions, providing
information on Regge factorisation and on proton structure.

This highly varied phenomenology provides an extraordinarily rich 
experimental laboratory for testing new theoretical ideas. We will see in
the following that the proposed THERA project is particularly well adapted
for tests of quantum chromodynamical descriptions of this high-energy domain.

\section{Lessons from HERA}

\begin{figure}
  \sidecaption
  \epsfig{file=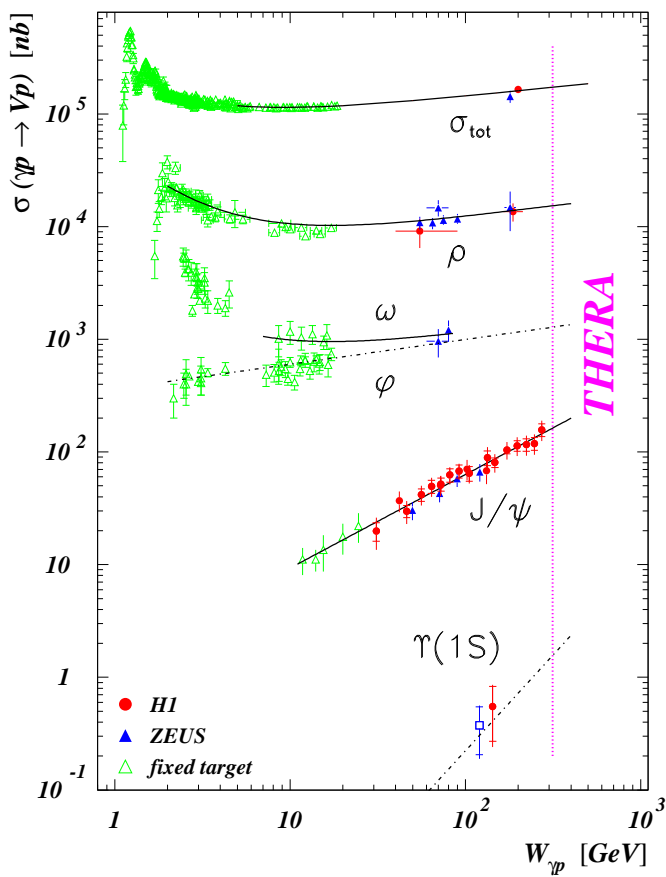,width=9.cm}
  \caption{
    Energy dependence of vector-meson photoproduction cross sections
for $\rho^0$, $\omega$, $\phi$, ${\rm J}/\psi$, and $\Upsilon$ mesons~\protect\cite{merkel}}
  \label{fig:vmfig}
\end{figure}

Experimental investigations of vector-meson production at 
HERA~\cite{pl_483_360,epj_6_603,epj_2_247,epj_13_371,*pl_483_23,*epj_10_373,*pl_421_385,*zfp_75_607,*np_472_3,*np_468_3,*np_463_3,*pl_338_507,*pl_487_273,*epj_12_393,*epj_14_213,*pl_437_432,*zfp_76_599,*zfp_75_215,*zfp_73_253,*pl_380_220,*pl_377_259,*zfp_73_73,*zfp_69_39,*pl_356_601,*pl_350_120}
(see~\cite{stmp_140} for a review) 
have provided a wide
variety of insights into the dynamics of both soft and hard diffractive
processes. The high flux of quasi-real photons from the electron beam
permitted detailed measurements of both elastic and proton-dissociative
photoproduction of $\rho^0$, $\omega$, $\phi$,  ${\rm J}/\psi$, and
$\Upsilon$ mesons. Power-law scaling with the photon-proton centre-of-mass
energy, $W_{\gamma p}$, was observed for the ${\rm J}/\psi$, as is illustrated by Fig.~\ref{fig:vmfig}.

\begin{figure}
  \sidecaption
  \epsfig{file=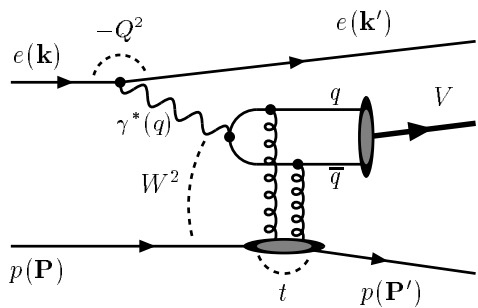,width=6cm}
\caption{ 
Schematic diagram illustrating exclusive vector-meson production as
mediated by the
exchange of a gluon pair in a colour-singlet state} 
\label{fig:pqcddiag}
\end{figure}

\begin{figure}[htbp]
\begin{minipage}{\textwidth}
\begin{center}
  \epsfig{file=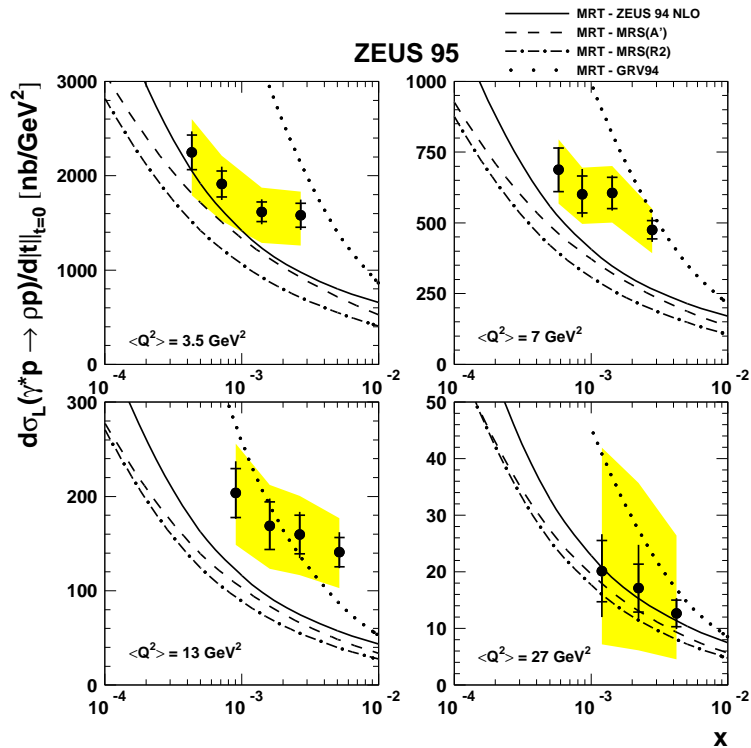,width=0.75\textwidth}
\end{center}
\end{minipage}
  \caption{
The measured forward longitudinal cross section, $\left. {\rm d}\sigma_{\rm L}^{\gamma^{\star}{\rm p}} /{\rm d}|t| \right| _{t=0}$, 
as a function of $x$
for $\rho^0$ electroproduction~\protect\cite{epj_6_603}.
The inner error bars represent statistical
uncertainties; the outer error bars indicate the quadratic sum of
statistical and systematic  uncertainties.
The shaded areas indicate additional normalisation uncertainties due to
the
proton dissociation background subtraction as well as the 
measured values of the 
$R=\sigma_{\rm L}^{\gamma^{\star}{\rm p}} / \sigma_{\rm T}^{\gamma^{\star}{\rm p}}$ ratio  and the $t$-slope parameter $b$.
The curves  show  the predictions by Martin, Ryskin and
Teubner~\protect\cite{pr_55_4329} and correspond to various
gluon parameterisations, indicated as follows: full lines -- 
ZEUS 94 NLO~\protect\cite{zfp_72_399},
dashed lines --  MRSA$^{\prime}$~\protect\cite{pl:b354:155}, dashed-dotted lines --  MRSR2~\protect\cite{pl:b387:419}, and dotted
lines --  GRV94~\protect\cite{zfp:c67:433}
}
  \label{fig:rhodsigdx}
\end{figure}

\begin{figure}
\begin{minipage}{\textwidth}
\begin{center}
  \epsfig{file=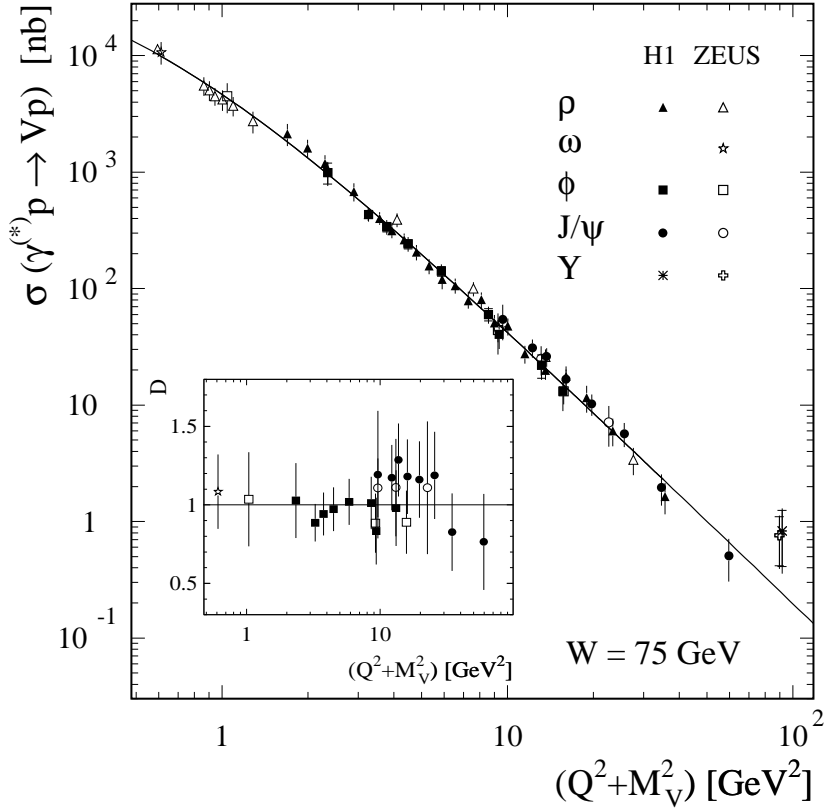,width=0.85\textwidth}
\end{center}
\end{minipage}
  \caption{
H1 and ZEUS measurements of the cross sections 
$\sigma (\gamma^* \rightarrow V p)$ as a function of 
(Q$^2+{\rm M}^2_{\rm V}$) for elastic $\rho^0$, $\omega$, $\phi$, J/$\psi$ 
and $\Upsilon$ production at the fixed value $W=75\, \gev$. 
The cross sections were scaled by SU(5) factors according to the 
quark charge content of the vector mesons. 
The error bars show 
statistical and systematic uncertainties added in quadrature. 
The curve corresponds to a fit to the H1 and ZEUS $\rho^0$ data, and 
the ratio, $D$, of the scaled $\omega$, $\phi$ and J/$\psi$ cross 
sections to this parameterisation is presented in the insert
}
  \label{fig:h1vmscale}
\end{figure}

\begin{figure}[htbp]
\vspace*{2mm}
\begin{minipage}[t]{\textwidth}
\begin{center}
\epsfig{file=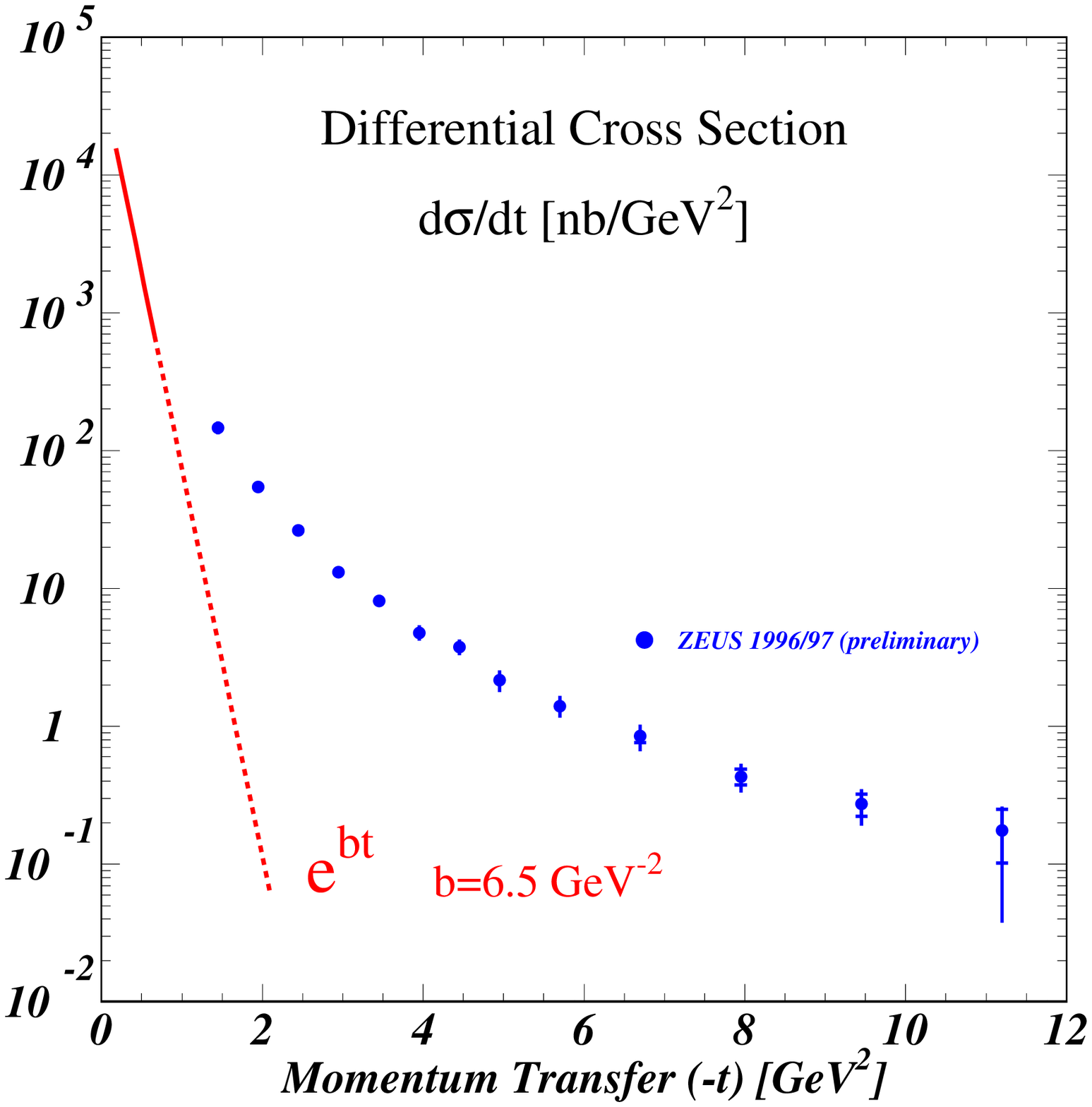, width=0.46\textwidth, bbllx=36, bblly=150, bburx=531, bbury=688, clip=}
\hfill
\epsfig{file=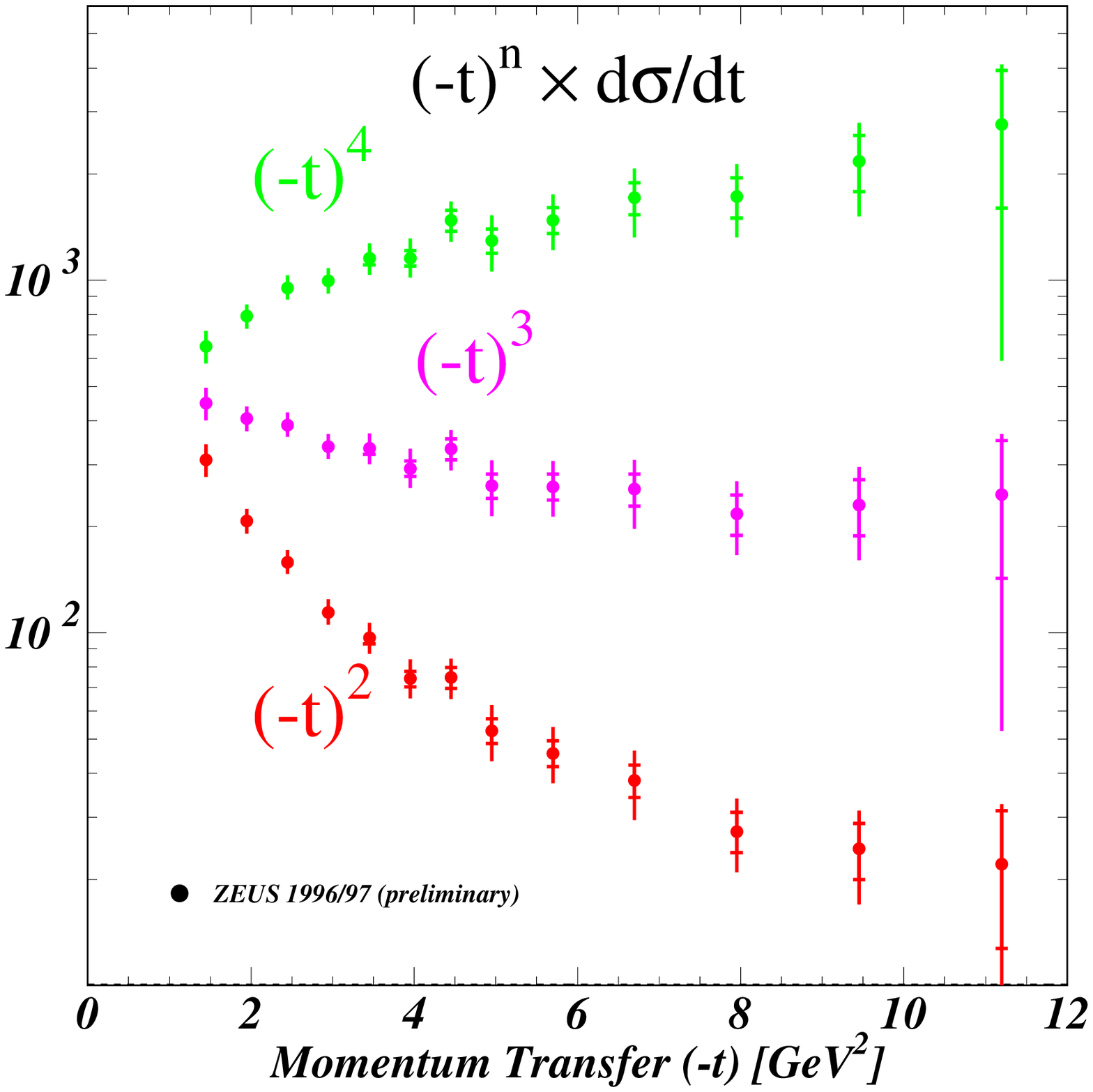, width=0.46\textwidth, bbllx=36, bblly=150, bburx=531, bbury=688, clip=}
\end{center}
\end{minipage}
\vspace*{5pt}
\caption{
\label{fig:rhoxsec}
a)~Differential cross section $\frac{{\rm d}\sigma}{{\rm d} t}$ for the process \mbox{$\gamma + p
\rightarrow {\rho^0} + Y$}, where $Y$ is a dissociated proton state. 
The line shows the $t$
dependence measured
for this process $\gamma + p \rightarrow {\rho^0} + Y$
($\propto e^{6.5\cdot t}$)~\protect\cite{epj_2_247} at low {$|t|$} (solid line) and its extrapolation
to higher {$|t|$} (dashed line) for comparison. 
b)~Differential cross section $\frac{{\rm d}\sigma}{{\rm d} t}$  multiplied by {(-t)}$^n$, where $n$=2, 3, 4
}
\end{figure}

It is instructive to compare the energy dependence of these 
vector-meson production cross sections to the photon-proton 
cross sections shown in Fig.~\ref{fig:sig}, 
where the steep energy dependence arises from the $x$ dependence of the 
gluon density in the proton.
The steep energy dependence measured for ${\rm J}/\psi$ mesons 
encouraged
a number of theoretical approaches based on perturbative 
QCD~\cite{zfp_57_89,*zfp_76_231,*pr_57_512}. Figure~\ref{fig:pqcddiag} 
shows a diagram 
illustrating this approach. Salient features of such calculations 
are an energy dependence
determined by the gluon density in the proton, flavour symmetry 
and the predominance of the longitudinal cross section 
for the light vector mesons at high momentum transfer.
The HERA measurement of $\Upsilon$ photoproduction 
resulted in theoretical 
investigations~\cite{pl_454_339,*jhep_99_02} which predict a very strong energy dependence
into the THERA region, with a large 
contribution from the off-diagonal gluon density. The phenomenological success
of these calculations supports the view that the factorisation scale can
be related to quark mass. Factorisation theorems have also been 
the object of theoretical investigations invoking the photon 
virtuality~\cite{pr_50_3134,pr_56_2982,pr_55_4329,*FKS,*pl_374_199}
and the momentum transferred to the proton~\cite{IKSS,*pr_53_3564,*pr_54_5523,*pl_449_306,*zfp_68_137,*pl_375_301} as  hard scales in
the production of vector mesons. These calculations demonstrated
remarkable sensitivity to the gluon density in the proton, 
since the forward cross sections
were shown to be proportional to its square.
Measurements at HERA 
of $\rho^0$, $\omega$, $\phi$ electroproduction and high-$t$ photoproduction
have served as testing grounds
for these calculations. Of particular interest is the experimental access
to the helicity structure of these processes via analysis of
decay-angle distributions, since these reflect not only the helicity selection rules but also meson structure~\cite{pr_58_114026,*np_545_505,*jetpl_69_294}.
Figure~\ref{fig:rhodsigdx} 
compares the ZEUS measurement of the 
differential cross section 
$\left. {\rm d}\sigma_{\rm L}^{\gamma^{\star}{\rm p}} /{\rm d}|t| \right| _{t=0}$ for $\rho^0$ production to
the results of QCD calculations~\cite{epj_6_603}, illustrating the degree
of consistency. 
The H1 collaboration has shown
a remarkably consistent
scaling behaviour common to the $\rho^0$, $\omega$, $\phi$, and J/$\psi$
mesons by plotting the production cross sections as a function
of Q$^2+{\rm M}^2_{\rm V}$, as shown in Fig.~\ref{fig:h1vmscale}~\cite{pl_483_360}.
This smooth behaviour is surprising from the point of view of the QCD
models, given that the helicity analyses have shown the relative contributions
of the longitudinal and transverse cross sections to depend strongly on Q$^2$,
and the QCD models predict very different scaling behaviour for these two
contributions. An investigation into high-$t$ $\rho^0$ and $\phi$
photoproduction by the ZEUS collaboration~\cite{ichep00_442}  has recently turned up another surprise. 
This first measurement of
vector-meson photoproduction at momentum transfers far exceeding the hadronic
confinement scale, extending 
into a region where power-law scaling is observed, 
permits an accurate determination of the power. It was found that the 
$\phi$/$\rho^0$ ratio reaches the SU(3) symmetric value in the same
region of momentum transfer where the power-law scaling takes over from
the exponential dependence observed at low $t$. The measurements of
the decay-angle distributions showed the vector mesons to be transverse.
This result, together with the extremely hard spectrum observed in the 
$t$~distribution (see Fig.~\ref{fig:rhoxsec})~\cite{jacaps}, appears to be at odds with the 
QCD helicity selection rules~\cite{IKSS,*pr_53_3564,*pr_54_5523,*pl_449_306,*zfp_68_137,*pl_375_301}.

These studies of exclusive vector-meson production have shown that the
transition region from the domain of applicability of perturbation theory
to the domain where long-distance strong-interaction dynamics
applies
can be scanned in photon virtuality and in
the momentum transfer at the proton vertex. Such measurements
have led to detailed theoretical consideration of the
interaction size scaling with energy~\cite{GG,pl_424_191}
and the relationship of diffraction to the mechanism of confinement~\cite{DD}.
\section{Requirements on the performance of machine and detector}

Measurements of 
vector-meson production at THERA will benefit not only from the
extended kinematic reach to low $x$ (high energy, wider rapidity range) and
high Q$^2$, but also from the improved coverage of the THERA detector
at small angles and from tagging systems in both the proton and electron flight
directions designed with forethought and the
benefit of the experience obtained at HERA. This experience has made
clear the importance of 
careful design and close interaction with the THERA machine group.
In the forward direction, the lack of high-$t$ acceptance in the proton
spectrometers prevents HERA studies of essential importance to QCD descriptions
of exclusive processes. A lack of instrumentation in the the region
of proton dissociation has resulted in the dominant source
of systematic uncertainties for the measurement of elastic 
cross sections being the subtraction of this background. At THERA, improvement
will come from requirements on the detector geometry independent of those
imposed by studies vector-meson production~\cite{theradetector}, but further 
instrumentation of the low-M$_{\rm Y}$ region must also be taken into careful
consideration. In the rear direction, a series of photoproduction taggers
with associated bending magnets to select off-beam-momentum electrons will
provide full coverage of the available range in $W$. The 
additional tracking coverage
in the rear direction required by the investigations of
inclusive processes at low-$x$ will enable precise measurements 
of the vector-meson decay products, and so permit accurate 
reconstruction of $t$ and and the
decay-angles at high $W$. This is particularly important for the light
vector mesons, since the hadronic decays used for their identification, 
together with 
the limited rear tracking in the H1 and ZEUS detectors result in
a limitation to the $W$ range of 
$W{\mathrel{\rlap{\lower2pt\hbox{\hskip1pt\small$\sim$}}
    \raise2pt\hbox{\small$<$}}}150\, {\gev}$.

The program of measurements described above
has been performed with an integrated luminosity
corresponding to  
that estimated for less than one year's  
running time at THERA. The weak energy dependence of the diffractive 
cross sections at low momentum transfer for the light vector mesons 
ensures a high data rate
during the early THERA running, making vector-meson production a
principal contribution to the early physics program, just as 
was the case at HERA. However, many of the HERA studies of
vector-meson production will remain statistics-limited. In particular,
multiply differential studies of 
the perturbative region of photon virtuality and momentum transfer to
the proton require high integrated luminosity. Another example of
investigations requiring stable accelerator performance at high luminosity
are those of diffractive J/$\psi$ and $\Upsilon$ photoproduction. Elastic 
electroproduction of $\Upsilon$ mesons, where effects of the 
off-diagonal parton densities are dominant, await THERA operation. 
The helicity analyses of vector-meson production benefit from
the longitudinal electron polarisation of the electron beam
at THERA, since the spin-density matrix elements arising from circular
photon polarisation are otherwise inaccessible~\cite{np_61_381}.
The high cross sections at low momentum transfer and the need for
high statistics in the kinematic region of applicability of perturbative
calculational techniques mean that the study of vector-meson production
will play an important r\^ole  in the THERA physics program 
beginning with the early 
turn-on stage of the machine and continuing throughout
the achievement of its full high-luminosity potential.

\section{Conclusions}

The proposed THERA accelerator complex is conceived in the interest of
extending the energy frontier in our understanding of electron-proton
interactions. Simple extrapolation from the experience gained during
the first nine years of HERA operation yield the reliable conclusion that
THERA will make essential contributions to our understanding of the
dynamics of strong interactions and, in particular, to the application
of Quantum Chromodynamics as a means to achieve this understanding.
The theoretical descriptions of the short-distance vacuum-exchange processes
under investigation at HERA remain in their infancy; the discovery potential
remains high. The parameters of the THERA machine directly address
limitations to the present investigations of vector-meson production at
HERA. The broad kinematic ranges in energy and momentum transfer
accessible to the experimental investigation of 
diffractive vector-meson production
ensure that such studies 
will make essential contributions to the THERA physics program throughout
the entire duration of its operation.

\section{Acknowledgements}
The author thanks B.~Surrow for providing Fig.~\ref{fig:sig}.
This work was supported by the Federal Ministry for Education
and Research of Germany.

\bibliographystyle{../bib/thera}
{\raggedright
\bibliography{../bib/theratdr,../bib/crittenden}
}

\end{document}